\documentclass[conference]{IEEEtran}
\usepackage{verbatim}
\usepackage{epsfig,bbm}
\usepackage[colorlinks,bookmarksopen,bookmarksnumbered,citecolor=red,urlcolor=red,]{hyperref}
\usepackage{CJK}
 \usepackage{indentfirst}
 \usepackage{multirow}
  \usepackage{epstopdf}
\usepackage{graphicx}
\usepackage{footmisc}
\graphicspath{{./figures/}}
\usepackage{amsfonts}
\usepackage{mathrsfs}
\usepackage{setspace}
\usepackage{amsmath}
 \usepackage{algorithm,algorithmic,amsbsy,amsmath,amssymb,epsfig,bbm,mathrsfs, bbm} 
\usepackage{amsthm}
\usepackage{verbatim}
\usepackage[subfigure]{graphfig}
 
\begin{document}

\title{ Payoff Allocation of Service Coalition in Wireless Mesh Network: A Cooperative Game Perspective}
\author{\\ Xiao Lu, Ping Wang, Dusit Niyato \\
   ~School of Computer Engineering, Nanyang Technological University, Singapore\\
  Email: \{Luxiao, Wangping, Dniyato\}@ntu.edu.sg
  }

\markboth{}{Shell \MakeLowercase{\textit{et al.}}: Bare Demo of
IEEEtran.cls for Journals}

\maketitle

\begin{abstract}

In wireless mesh network (WMN), multiple service providers (SPs) can cooperate to share resources (e.g., relay nodes and spectrum), to serve their collective subscribed customers for better service. As a reward, SPs are able to achieve more individual benefits, i.e., increased revenue or decreased cost, through efficient utilization of shared network resources. However, this cooperation can be realized only if fair allocation of aggregated payoff, which is the sum of the payoff of all the cooperative SPs, can be achieved. We first formulate such cooperation as a coalitional game with transferable utility, specifically, a linear programming game, in which, each SP should obtain the fair share of the aggregated payoff. Then we study the problem of allocating aggregated payoff which leads to stable service coalition of SPs in WMN based on the concepts of dual payoff and Shapley value. 

\end{abstract}

\IEEEpeerreviewmaketitle
\section{Introduction}

In wireless mesh network (WMN), the utilization of available resource, i.e., relay node and spectrum, and the cost of routing, i.e., consumption of the network, can be substantially improved and reduced, respectively, through cooperation. That is, multiple service providers (SPs) may form a coalition to pool their resources to serve all their subscribed customers together, which leads to higher satisfied service~\cite{Lin10}.

Coalition of WMN benefits SPs in the following two ways. First, SPs coalition may substantially lead to higher revenue. In some WMNs, relay nodes (e.g., access points and routers) are deployed dispersedly. For each SP, if there is no cooperation, the limited number of nodes and their dispersive locations largely confine the link capacity that can be provided for flow transmission. Cooperation among SPs could increase the number of available relay nodes for each SP which in turn improves available link capacity. Accordingly, a larger amount of flow rate requirement can be satisfied by efficiently utilization of resources in the WMN, which leads to higher aggregated revenue. Second, SPs coalition may lead to lower network cost. A cost (e.g., due to the usage of energy and spectrum) incurs when a node transmits or relays traffic flow. Under non-cooperation, an SP can only use its own relay nodes for flow transmission which largely constrains the routing. SPs cooperation could increase the number of multi-hop routes which in turn provides more options of packet forwarding for flows. If the optimal route from source to destination node under cooperation involves less relay nodes than under non-cooperation, the cost for flow transmission is reduced. 

\begin{figure}[t]
\centering
\includegraphics[width=0.5\textwidth]{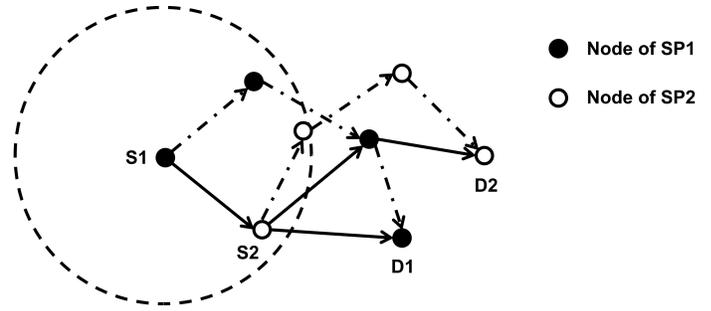}
\caption{An example of SPs cooperation in WMN.} \label{example}
\end{figure}

Fig. \ref{example} shows an example of two SPs cooperation. The dashed circle line represents the transmission range of the source node S$1$. Due to the limited transmission range of each node, when there is no cooperation, the three-hop route represented as the semi-dashed arrow lines is the only route available for SP$1$ to transmit flow from source node S$1$ to destination node D$1$. This route incurs the cost of three nodes for transmission. Cooperation between the two SPs can reduce the cost as only two nodes are used for the transmission of this flow with the route represented as the solid arrow lines. Similarly, this is also the case with SP$2$ for flow transmission from S$2$ to D$2$. Besides, under cooperation, multiple routes can be utilized together to support a higher rate requirement that a single route cannot satisfy, which leads to better service for customers and higher revenue for SPs.

Although SPs coalition brings obvious benefits, a major challenge arise in the formation of coalition. 
How the aggregated payoff (i.e., revenue) shall be allocated to each SP so that the coalition of them can be stabilized (i.e., none of them has incentive to leave). In this paper, we first model the SPs cooperation as a coalitional game with transferable utility~\cite{Driessen10}. An optimization problem based on linear programming game is formulated. Then, to address the aforementioned problem, we focus on the investigation of payoff allocation solutions which always lead to cooperation among SPs.

The importances of payoff allocation solution in WMN which motivate our study are $1$) from an individual point of view, it is efficient to let each SP in the service coalition achieve a fair sharing according to some common recognized principles; $2$) from a social point of view, it encourages formation of service coalition which provides improved network service for customers; and $3$) from a commercial point of view, it helps each SP to form the stable coalition which helps each one to gain. The analysis of cooperation behavior and payoff allocation solutions are useful for SPs in WMN especially from the economic perspective.

\section{Related Works}
Recently, many research interests have been focused on the allocation problem of joint resource sharing in WMN~\cite{Guha07}. 
Coalitional game theoretical framework, which is also adopted in this paper, was widely used to model the network cooperation. The existing approaches are mainly based on the concept of Shapley value~\cite{Shapley53}.~\cite{Richard10} presented a clean-state Shapley payoff allocation solution, under which Internet service providers have incentive to reach an equilibrium that maximizes both individual profit and the system's social welfare by performing globally optimal routing decisions. In~\cite{Maac}, with a similar system model to that in~\cite{Richard10}, the Shapley value solution was used under structured topologies, and a dynamic programming procedure to calculate the Shapley value solution was developed for general topologies. Similarly, profit sharing scheme based on Shapley value was also exploited in~\cite{Le08} for radio access in cooperative networks,~\cite{Pan09} for spectrum auction in wireless network, and~\cite{Niyato06} for heterogeneous wireless assess in 4G networks.

\section{Network Model and Assumptions}

\subsection{Wireless Mesh Network Model}
We consider a WMN consisting of nodes belonging to multiple SPs. Let $\mathcal{M}=\{ 1, 2,\ldots,M\}$ denote the set of SPs, $\mathcal{N}^{(m)}=\{ 1, 2,\ldots,N_{m}\}$ the set of nodes of SP$m$, and $N_{m}$ the total number of nodes belonging to SP$m$, $m \in \mathcal{M}$. In the SPs coalition, each node $i \in \mathcal{N}^{(m)}$ not only needs to support internal flow transmission demand, but also serves as a relay node for other cooperative SPs. In a WMN, multi-hop relaying is usually used to route flow session from source node to destination node. For optimality, we assume that each flow session can be split for multi-path routing.
Let $b_{ij}$ denote the frequency band assigned to link $(i, j)$, and $W(b_{ij})$ denote the bandwidth of frequency band $b_{ij}$. 
Let $\mathcal{T}_{i}$ denote the set of nodes that are within the transmission range of node $i $. 
$\mathcal{B}=\bigcup_{i \in \mathcal{N}, j \in \mathcal{T}_{i}} \{ b_{ij} \}$ represents the set of all assigned frequency bands in the network. We assume that each SP owns sufficient frequency bands and each link $(i, j)$ between two nodes $i $ and $j \in \mathcal{T}_{i}$ is allocated with one frequency band at the initial stage of WMN. To avoid channel interference, there is no reuse of the same band within the interference range of the nodes in WMN.

Let $f_{(i, j)}(l^{(m)})$ denote the data rate on link $(i, j)$ attributed to a flow session $l^{(m)}$, and $\mathcal{L}^{(m)}$ denote the set of flow sessions of SP$m$. Since, in WMN, a flow session from a source node may traverse through multiple relay nodes to reach its destination node, we consider the following two cases.

1) Let $s(l^{(m)})$ and $d(l^{(m)})$ denote the source node and the destination nodes of flow $l^{(m)}$, respectively, and $\mathcal{I}^{(m)}$ and $\mathcal{D}^{(m)}$ the set of source nodes and the set of destination nodes of flow sessions owned by SP$m$, respectively. If node $i$ is the source or destination nodes of flow session $l^{(m)}$, i.e., $i=s(l^{(m)})$ or $i=d(l^{(m)})$, then
\begin{eqnarray} 
\sum_{j \in \mathcal{T}_{i}} f_{(i, j)}(l^{(m)})=r(l^{(m)}) \label{source} \hspace{2mm} \text{or} \\
 \sum_{p \in \mathcal{T}_{i}} f_{(p, i)}(l^{(m)})=r(l^{(m)}), \label{destination}
\end{eqnarray}
where $r(l^{(m)})$ is the aggregated rate of flow session $l^{(m)}$.

2) If node $i$ is an intermediate relay node of flow session $l^{(m)}$, i.e., $i \neq s(l^{(m)})$ and $i \neq d(l^{(m)})$, then
\begin{eqnarray} 
\sum^{j \neq s(l^{(m)})}_{j \in \mathcal{T}_{i}} f_{(i, j)}(l^{(m)})=\sum^{p \neq d(l^{(m)})}_{p \in \mathcal{T}_{i}} f_{(p, i)}(l^{(m)}). \label{intermediate}
\end{eqnarray}
It is clear that if (\ref{source}) and (\ref{intermediate}) hold, then (\ref{destination}) must be satisfied. Therefore, it is sufficient to have only (\ref{source}) and (\ref{intermediate}) in the formulation. 

Let $c_{ij}$ denote the capacity of link $(i ,j)$. The aggregated data rate on each link $(i,j)$ cannot exceed the link's capacity. Thus, we have the following constraint,
\begin{eqnarray} 
\label{rate} \sum^{i \neq d(l^{(m)}), j \neq s(l^{(m)})}_{l^{(m)}\in \mathcal{L}^{(m)}}f_{(i, j)}(l^{(m)}) \leq c_{ij}.
\end{eqnarray}

Let $f^{\star}_{(i, j)}(l^{(m)})$ denote the maximal rate that a flow session $l^{(m)}$ can achieve on link $(i, j)$, confined by (\ref{source}), (\ref{intermediate}) and (\ref{rate}), and $F^{\star}(l^{(m)})$ denote the maximal aggregated rate of a flow session $l^{(m)}$. We have $F^{\star}(l^{(m)})= \sum^{i = s(l^{(m)})}_{j \in \mathcal{T}_{i}} f^{\star}_{(i, j)}(l^{(m)})$. The aggregated rate of flow session $l^{(m)} \in \mathcal{L}^{(m)}$ is constrained by
\begin{eqnarray}
r(l^{(m)})= \left\{
\begin{array}{rcl}
R(l^{(m)}) , & & { R(l^{(m)})  \leq F^{\star}(l^{(m)}) }, \\
 F^{\star}(l^{(m)}) , & & \text{otherwise},
\end{array} \right. \end{eqnarray}
where $R(l^{(m)})$ is the transmission rate requirement of $l^{(m)}$.

\section{Optimization Formation of SPs Coalition in WMN}
\begin{figure*} 
\normalsize
\begin{eqnarray}
 \max_{f_{(i, j)}(l^{(m)})}: & \quad v(\mathcal{S})=\sum_{m \in \mathcal{S} }\sum_{l^{(m)} \in \mathcal{L}^{(m)}} \sum^{i = s(l^{(m)})}_{j \in \mathcal{T}_{i}} f_{(i, j)}(l^{(m)}) P - \sum_{m \in \mathcal{S}} \sum_{l^{(m)} \in \mathcal{L}^{(m)}} \sum_{(i, j) \in \Pi(l^{(m)}) } f_{(i, j)}(l^{(m)}) C \label{lpg} \\ 
\mbox{s.t.}: \hspace{5mm} & \sum_{j \in \mathcal{T}_{i}} f_{(i, j)}(l^{(m)})- r(l^{(m)})=0, \hspace{5mm} ( i \in \mathcal{I}^{(m)} , l^{(m)} \in \mathcal{L}^{(m)}, m \in \mathcal{S}), \label{one} \\%
& \sum_{m \in \mathcal{S}} \sum_{l^{(m)} \in \mathcal{L}^{(m)}} \sum^{ j \neq s(l^{(m)})}_{j \in \mathcal{T}_{i}} f_{(i, j)}(l^{(m)}) - \sum_{m \in \mathcal{S}} \sum_{l^{(m)} \in \mathcal{L}^{(m)}} \sum^{ p \neq d(l^{(m)})}_{p \in \mathcal{T}_{i}}f_{(p, i)}(l^{(m)}) = 0, \nonumber \\ 
&(i \in \mathcal{N}^{(m)} \setminus \{\{\mathcal{I}^{(m)}\} \cup \{ \mathcal{D}^{(m)} \}\} ), \label{two} \\
& f_{(i, j)}(l^{(m)}) \geq 0, P > 0, C>0, \hspace{5mm} (l^{(m)} \in \mathcal{L}^{(m)}, i \in \mathcal{N}^{(m)}, m \in \mathcal{S}, i \neq d(l^{(m)}), j \in \mathcal{T}_{i}, j \neq s(l^{(m)}) ). \label{three} 
\end{eqnarray}
\vspace{1pt} \hrulefill
\end{figure*}

In this section, we develop a linear programming model for the payoff optimization problem of cooperative SPs in WMN. 
Coalition game with transferable utility~\cite{Driessen10}, which allows side payment among SPs, is adopted to model the SPs cooperation.

From a global perspective, we wish that, through SPs cooperation, all the flows would choose feasible routings that maximize the aggregated payoff of the entire network. This SPs coalitional game in characteristic form is $(\mathcal{M},v(\mathcal{S}))$ for $\mathcal{S} \subseteq \mathcal{M}$. This SPs coalitional game in characteristic form is $(\mathcal{M},v(\mathcal{S}))$ for $\mathcal{S} \subseteq \mathcal{M}$, where $\mathcal{M}$ is the set of players and $v(\mathcal{S})$ is the maximum aggregated payoff available for division in any arbitrary way among the members of $\mathcal{S}$. 
Let $P$ denote the normalized revenue per unit data rate provided, and $C$ denote the normalized cost per unit data rate transmitted by source node or forwarded by relay node due to energy consumption and spectrum usage. Usually, for the commercial reason, it is common to assume $P >> C$. The optimization problem for SPs coalition is to maximize their aggregated payoff, i.e., revenue collected for providing service subtracts the cost of using network nodes. Putting all the constraints for routing aforementioned in Section III, we can formulate the payoff optimization of coalition $\mathcal{S}$ as the linear programming problem defined in (\ref{lpg}), subjected to (\ref{one}) (\ref{two}) and (\ref{three}), where $\Pi(l^{(m)})$ is the set of links that are in use for the transmission of flow session $l^{(m)}$.

The optimization problem (\ref{lpg}) provides the maximum aggregated payoff of SPs in a coalition $\mathcal{S}$.
It is straightforward to show that, for any disjoint coalition $\mathcal{S} \subseteq \mathcal{M}$ and $ \mathcal{T} \subseteq \mathcal{M}$, $v(\mathcal{M}) \geq v(\mathcal{S}) + v(\mathcal{T})$. In other words, the coalitional game is \emph{super-additive}. That is, the grand coalition of all SPs attains the maximum possible aggregated payoff. 

\newtheorem{definition}{Definition}
\begin{definition}
A coalitional game $(\mathcal{M}$, $v(\cdot))$ is called a linear programming game if there exists an $m \times p$ matrix $A$, an $m \times r$ matrix $H$, and vectors $g(\mathcal{S})\in \mathbb{R}^{p}$ and $t(\mathcal{S})\in \mathbb{R}^{r}$ for all $\mathcal{S} \in 2^{\mathcal{M}} \setminus \{\phi \}$, where $\{\phi\}$ represents empty set, such that for the optimization problem,
\begin{align}
v(\mathcal{S}): & \max_q \quad c \cdot q \nonumber \\ 
\mbox{s.t.} \hspace{3mm}
& qA \leq g, \quad  qH=t, q \geq 0. \nonumber 
\end{align}
 $v(\mathcal{S}) = v_{p}(A,H,g(\mathcal{S}),t(\mathcal{S}),c)$ holds.
\end{definition}

It is obvious that  (\ref{lpg}) satisfies all the requirements of Definition 2. Thus, (\ref{lpg}) is a linear programming game.

\section{Payoff Allocation Solution}
In this section, the concept of stability in coalitional game is introduced. We then introduce two solution approaches of payoff allocation in the coalitional game theory based on the concepts of \emph{dual payoff} and \emph{Shapley value}.

\subsection{Core of Coalitional Game}

How to divide the aggregated payoff among the cooperative players in coalitional game is the key for a stable coalition. According to~\cite{Driessen10}, \emph{core} is the set of feasible payoff vectors for players in the grand coalition that none of player has incentive to leave, which is analogous to the idea behind a \emph{Nash equilibrium} of a non-cooperative game.
\begin{definition}
A real valued vector $\textbf{x}=(x_{i}, i\in\mathcal{M})$ is said to be an imputation if $\sum_{i\in\mathcal{M}}x_{i} = v(\mathcal{M})$ and $x_{i} \geq v(\{i\})$, $\forall i \in \mathcal{M}$. 
The core of the coalitional game with transferable payoff $(\mathcal{M}, v(\cdot))$ is the set of all imputations \textbf{x} for which $\sum_{i \in \mathcal{S}} x_{i} \geq v(\mathcal{S})$, $\forall \mathcal{S} \subseteq \mathcal{M}$. In other words, the core can be represented as follows:
\begin{eqnarray} 
\mathcal{C}= \{ x \in \mathbb{R}^{\mathcal{M}} |\sum_{i \in \mathcal{M}} x_{i} =v(\mathcal{M}), \sum_{i \in \mathcal{S}} x_{i} \geq v(\mathcal{S}), \forall \mathcal{S} \subseteq \mathcal{M} \}. \nonumber
\end{eqnarray}
\end{definition}

In the proposed SPs coalitional game, any reasonable bases for allocating the aggregated payoff need to be imperative to motivate the SPs to join the grand coalition which maximizes the aggregated payoff. In other words, any set of allocated payoffs to the SPs should lie in the core of the game. 
Since the SPs coalitional game defined in (\ref{lpg}) is super-additive, the core always exists~\cite{Driessen10}. The core may not be a unique imputation and the set of imputations in the core may be quite large for a coalitional game that is super-additive. Any imputation in the core is stable; however, this imputation is only desirable if it is obtained according to some principles that can guarantee fairness and uniqueness.
In the following, we address the fair payoff allocation problem in SPs coalitional game by adopting two well-known concepts from cooperative game theory, i.e., the dual payoff and Shapley value solutions.

\subsection{Dual Payoff Solution}

To every linear programming problem, there is a dual problem that is intimately connected to the original one.
The dual optimal solutions for the optimization problem of linear programming game can be obtained by solving the dual problem of deterministic equivalent linear programming of payoff allocation~\cite{Driessen10}. A linear programming game and its dual problem are equivalent in some sense, e.g., the cores of a game and its dual are the same. The dual problem of SPs coalitional game defined by (\ref{lpg}) can be expressed as (\ref{dual}),
where $y_{i}$, $z_{i}$, $e_{(i, j)}(l^{(m)})$ are the dual variables corresponding to the constraints (\ref{one}), (\ref{two}), and (\ref{three}), respectively.

\begin{figure*} 
\normalsize
\begin{eqnarray}
   \min_{y_{i}, z_{i}, e_{(i, j)}(l^{(m)})}: & \quad \sum_{m\in \mathcal{M}} \sum_{l^{(m)} \in \mathcal{L}^{(m)}} \sum_{(i, j) \in \Pi(l^{(m)})} e_{(i, j)}(l^{(m)}) \cdot r(l^{(m)}) -  \sum_{m\in \mathcal{M}} \sum_{i \in \mathcal{I}^{(m)}} y_{i} \cdot r(l^{(m)}) 
\label{dual} \\
\mbox{s.t.}: \hspace{5mm} &  e_{(i, j)}(l^{(m)}) + z_{i} + y_{i} = C-P, \hspace{5mm} (i \in \mathcal{I}^{(m)}, j \in \mathcal{T}_{i} ),\nonumber \\
& e_{(i, j)}(l^{(m)}) + z_{i} = C, \hspace{5mm} (i \in \mathcal{N}^{(m)} \setminus \{ \mathcal{I}^{(m)} \} ),\nonumber \\
& e_{(i, j)}(l^{(m)}) - z_{i} = C, \hspace{5mm} (i \in \mathcal{N}^{(m)} \setminus \{ \mathcal{D}^{(m)} \} ),\nonumber \\
& e_{(i, j)}(l^{(m)}) \geq 0, z_{i} \in \mathbb{R}, y_{i} \in \mathbb{R}, \hspace{5mm} (i \in \mathcal{N}^{(m)}, j \in \mathcal{T}_{i}, l^{(m)} \in \mathcal{L}^{(m)}, m \in \mathcal{M} ) . \nonumber 
\end{eqnarray}
\vspace{1pt} \hrulefill
\end{figure*}

The payoff allocation of SP$m$ in the grand coalition can be expressed as follows:
\begin{eqnarray} 
\mu_{m}(v(\mathcal{M}))= \sum_{l^{(m)} \in \mathcal{L}^{(m)}} \sum_{(i, j) \in \Pi(l^{(m)})} e_{(i, j)}(l^{(m)}) f_{(i, j)}(l^{(m)}) \nonumber \\ 
-  \sum_{i \in \mathcal{I}^{(m)}} y_{i} \cdot r(l^{(m)}) .
\end{eqnarray}
 
The \emph{dual payoff} is the payoff vectors $\{ \mu_{1}(v(\mathcal{M})), \ldots,$ $ \mu_{m}(v(\mathcal{M})) \}$ generated by the solution of dual problem. Two important properties of dual payoff solution, making it suitable for payoff allocation in SPs coalitional game, are the efficiency and rationality. For efficiency, the sum of the allocated payoffs of all SPs equals to the maximum aggregated payoff, i.e., $\sum_{m \in \mathcal{M}} \mu_{m}(v(\mathcal{M})) = v(\mathcal{M})$. For rationality, the allocated payoffs of all SPs must be more than or equal to those of sub-coalitions, i.e., $\sum_{m \in \mathcal{S}} \mu_{m}(v(\mathcal{M})) \geq v(\mathcal{S}), \forall \mathcal{S} \subseteq \mathcal{M} $.

\subsection{Shapley Value Solution}

\begin{definition}
 The marginal contribution of player $m$ to a set
$\mathcal{S} \subseteq \mathcal{M} \setminus \{m\}$ is defined as follows:
\begin{eqnarray} 
\triangle_{m} (v(\cdot), \mathcal{S}) = v(\mathcal{S} \cup \{m\})-v(\mathcal{S}).
\label{mc}
\end{eqnarray}
\end{definition}

Shapley value, first introduced in~\cite{Shapley53}, is a unique value based on the marginal contribution of each player to the coalition.

\begin{definition}
The Shapley value is obtained from 
\begin{eqnarray} 
\varphi_{m}(v(\mathcal{M}))=\sum_{\mathcal{S} \subseteq \mathcal{M} \setminus \{ m \}}\frac{|\mathcal{S}|!(|\mathcal{M}|-|\mathcal{S}|-1)!}{|\mathcal{M}|!} \triangle_{m}(v(\cdot), \mathcal{S}) .
\label{sv}
\end{eqnarray}
\end{definition}

The Shapley value is suitable for payoff allocation in SPs coalitional game due to the efficiency and individual fairness. The efficiency, as aforementioned, is the sum of the allocated payoffs for all SPs equals to the maximum aggregated payoff, i.e., $\sum_{m \in \mathcal{M}}\varphi_{m}(v(\mathcal{M}))=v(\mathcal{M})$. For individual fairness, Shapley value guarantees the allocated payoff of each SP to be more than or equal to the value of the individual SP, i.e., $\varphi_{m}(v(\mathcal{M})) \geq v(\{m\})$, for all $m \in \mathcal{M}$. Other properties of Shapley value, i.e., symmetry, uniqueness, dummy, strong monotonicity, and the details can be found in~\cite{winter02}.

In the SPs coalitional game, the Shapley value of each SP can be derived from the combination of (\ref{lpg}), (\ref{mc}) and (\ref{sv}).

\begin{figure*}
\vspace{-5mm}
 \graphfile*[30]{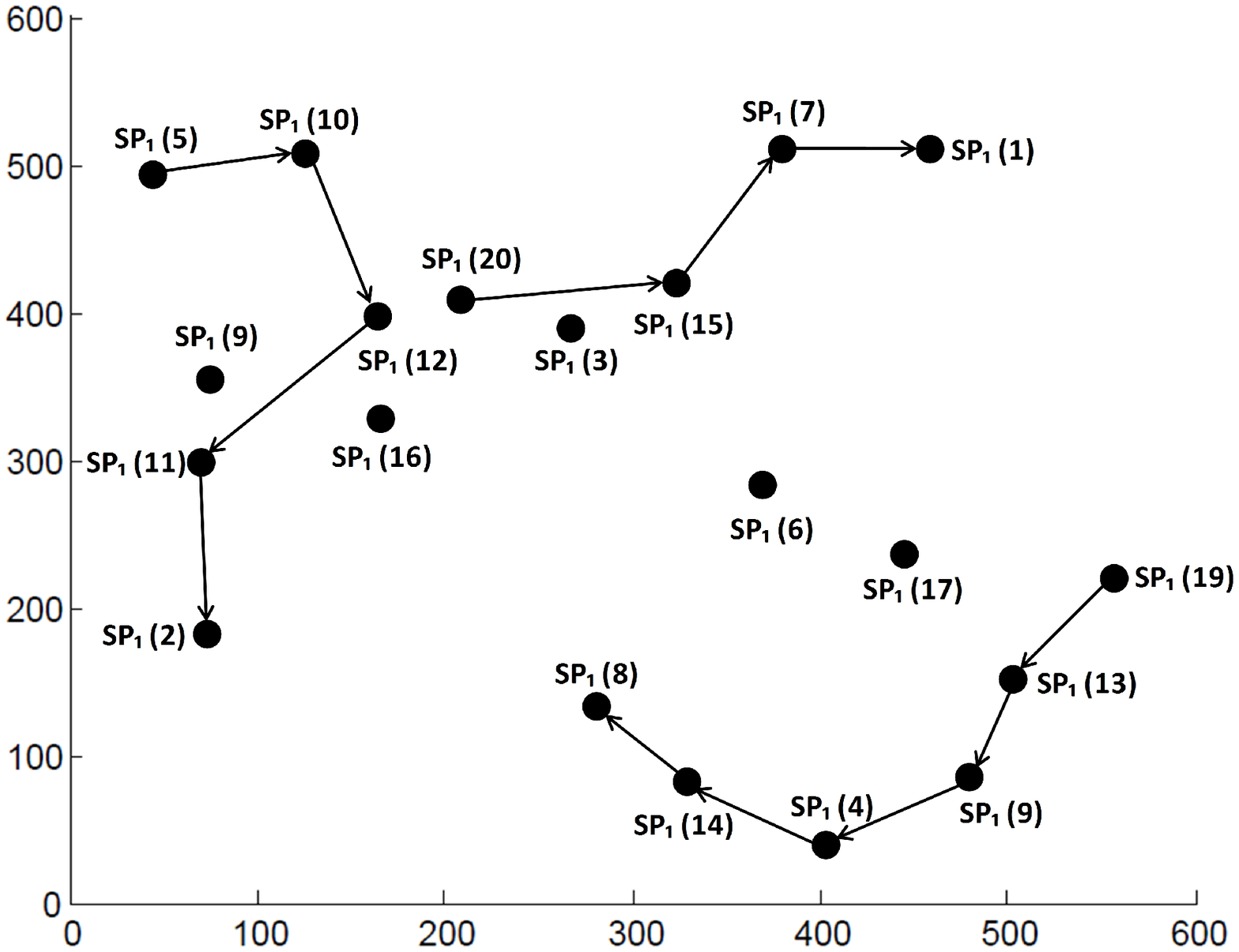}[Flow Sessions of SP$1$] 
 \graphfile*[30]{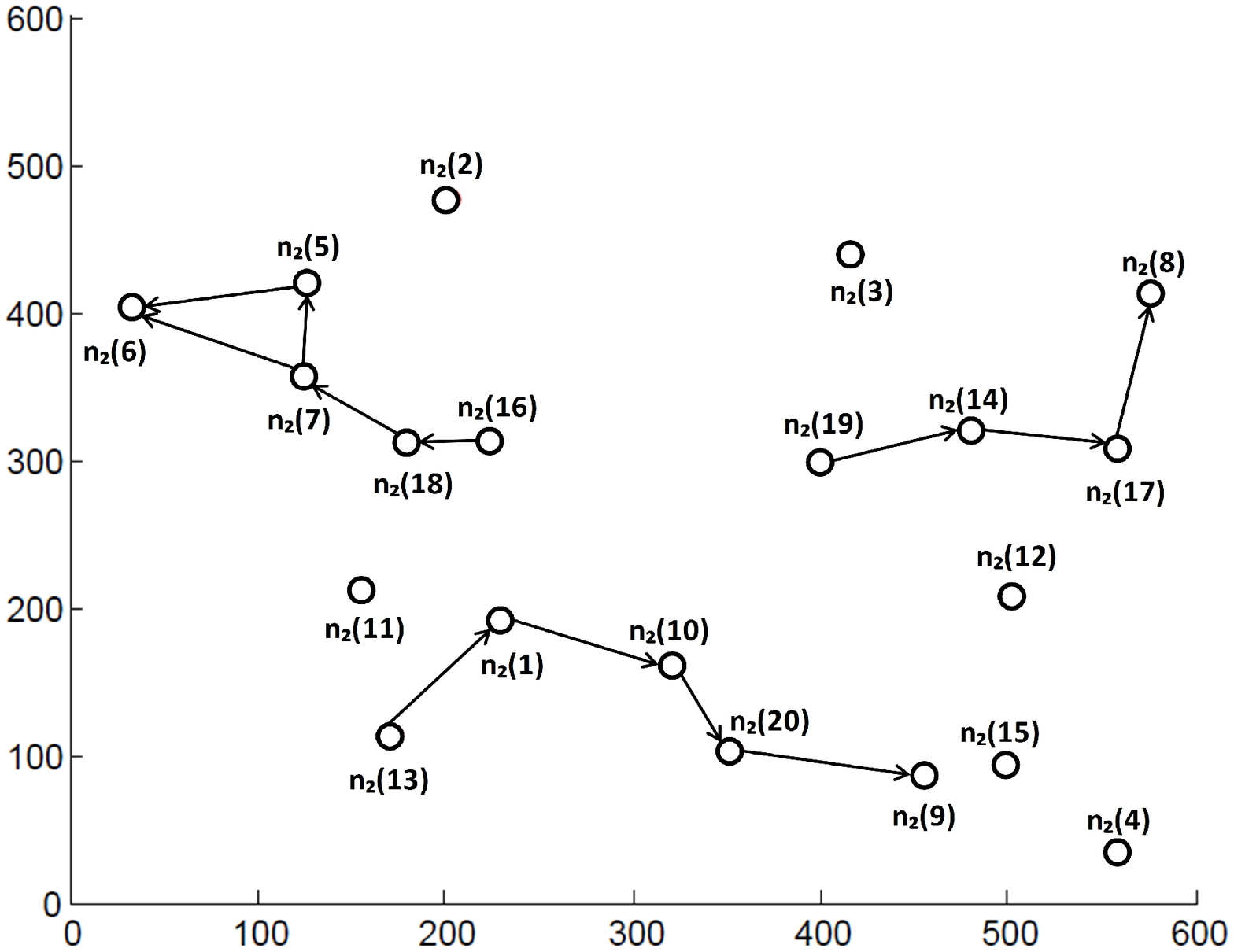}[Flow Sessions of SP$2$] \\
 \graphfile*[30]{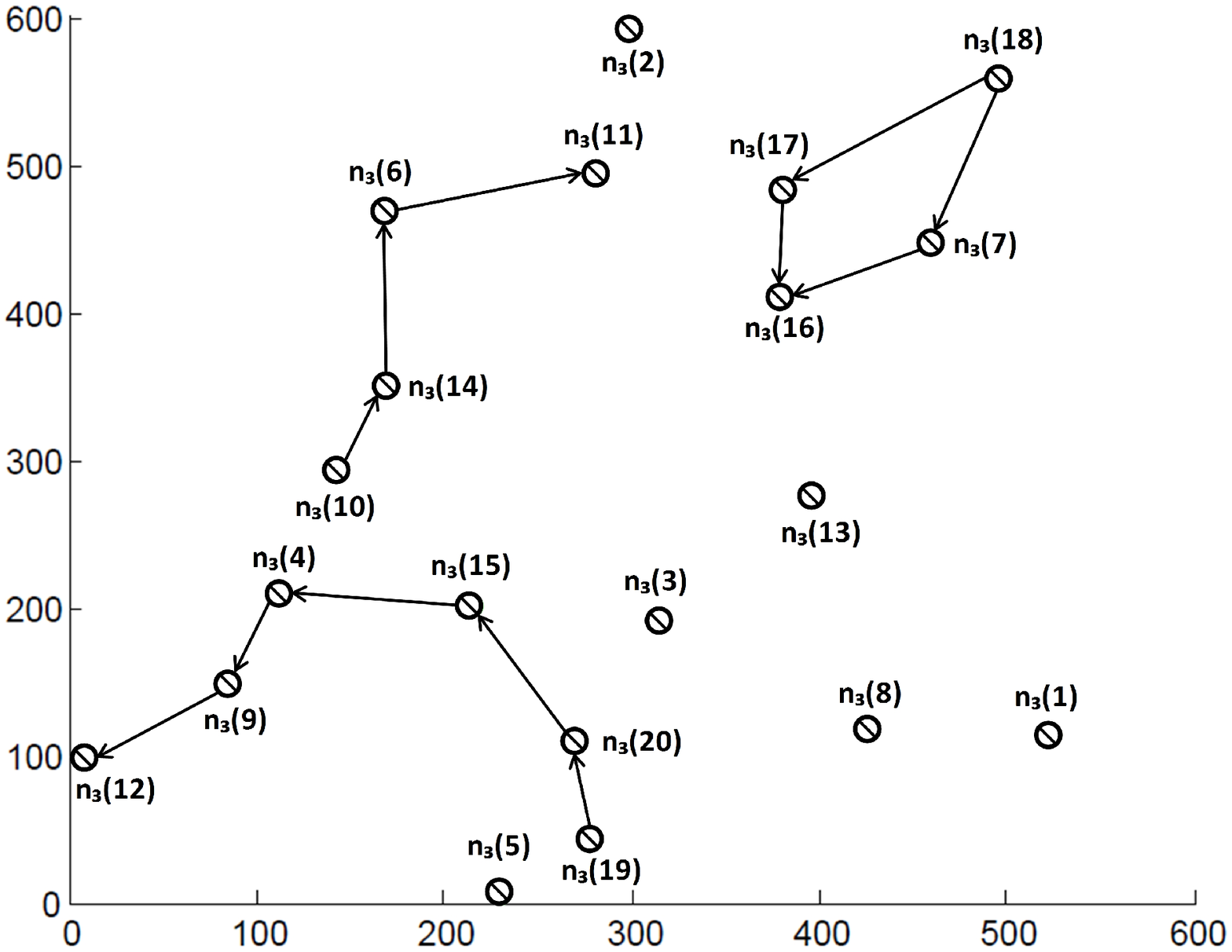}[Flow Sessions of SP$3$] 
 \graphfile*[30]{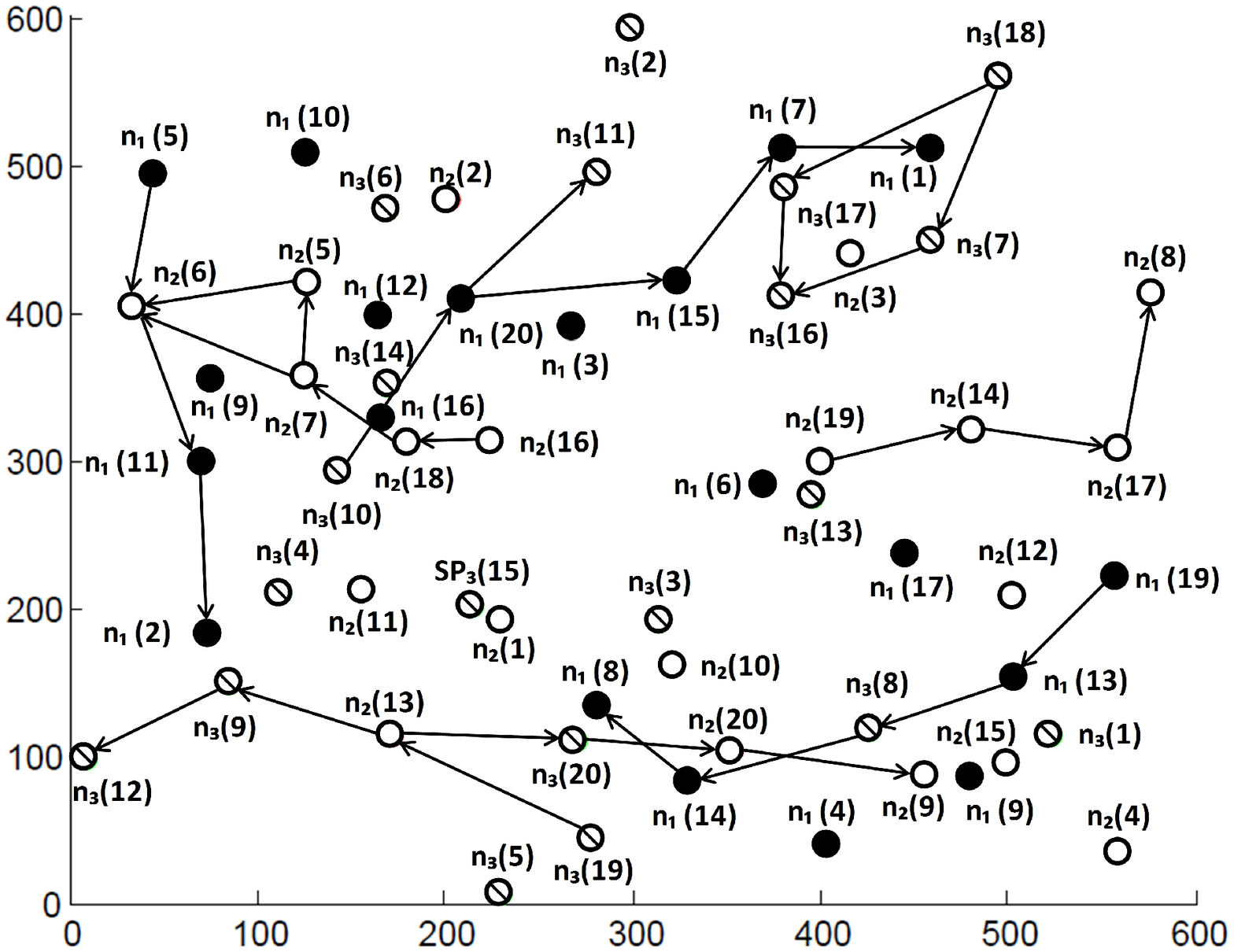}[Flow Sessions of Three-SP Coalition] 
 \caption{Network topology and flow sessions of three-SP coalition } \label{Network_topology} 
\end{figure*}

\begin{table}
\centering
\caption{Profile of flow sessions in three-SP network.} \label{Session} 
\begin{tabular}{|c|c|c|c|} 
\hline
Session & Source & Destination & Rate Req. (Kbps) \\ \hline
\hline
$l^{1}_{1}$ & $n_{1}(5)$ & $n_{1}(2)$ & $33$ \\
$l^{1}_{2}$ & $n_{1}(20)$ & $n_{1}(1)$ & $42$ \\
$l^{1}_{3}$ & $n_{1}(19)$ & $n_{1}(8)$ & $55$ \\ 
$l^{2}_{1}$ & $n_{2}(16)$ & $n_{2}(6)$ & $71$ \\
$l^{2}_{2}$ & $n_{2}(13)$ & $n_{2}(9)$ & $48$ \\
$l^{2}_{3}$ & $n_{2}(19)$ & $n_{2}(8)$ & $53$ \\
$l^{3}_{1}$ & $n_{3}(19)$ & $n_{3}(12)$ & $41$ \\
$l^{3}_{2}$ & $n_{3}(10)$ & $n_{3}(11)$ & $37$ \\
$l^{3}_{3}$ & $n_{3}(18)$ & $n_{3}(16)$ & $64$ \\
\hline 
\end{tabular}
\end{table}

\section{Numerical Results}

In this section, we present the numerical results of the two aforementioned payoff allocation solutions. The fairness and stability of the solutions are demonstrated by comparison in numerical simulations.
 
\subsection{Simulation Setting}
We consider a WMN consisting of three SPs, i.e., $\mathcal{M}=\{$SP$1$, SP$2$, SP$3\}$. Each SP has $20$ nodes randomly locate in a $600m \times 600m$ area. The bandwidth of each frequency band is set to be $W(b_{ij}) = 200$KHz ,  $\forall  b_{ij} \in \mathcal{B} $. The transmission range of each node is set to be $150m$. As for channel quality, 
we consider power propagation gain $g_{ij}=62.5 \cdot d^{-4}_{ij}$ like adopted in \cite{hou08}, where $d_{ij}$ is the length of link $(i, j)$. The capacity of each link is calculated according to Shannon Theorem.

For each SP, there are $3$ internal flow sessions. For each flow session, the source node and destination node are randomly selected and the rate requirement is randomly generated within $[20,80]$ Kbps. We assume that each SP charges its customers for flow service at the same price level and the service fee for providing a unit flow rate (i.e., 1 Kbps) is $10$. The cost for each node to transmit or forward is $1$ \emph{per} unit flow rate. Given all the network and flow session profiles in the entire network, the aggregated revenue is a constant. However, the coalition provides more flexible routing options which could reduce the cost and thus enhance payoff objective function (\ref{lpg}).

\subsection{Three-SP Coalition}

\begin{figure}[t]
\centering
\includegraphics[width=0.50\textwidth]{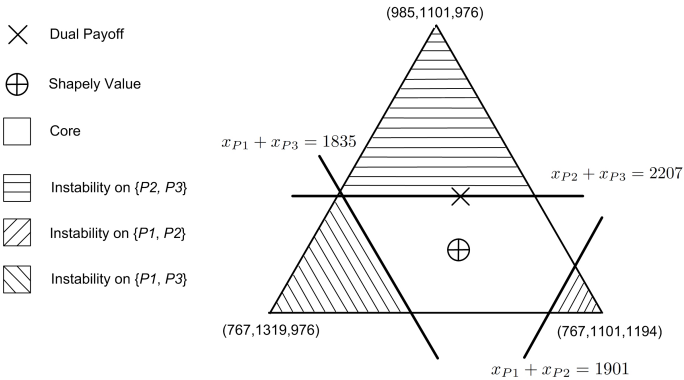}
\caption{Barycentric coordinates of the core, dual payoff and Shapley value for the numerical example.} \label{Barycentric}
\end{figure}

This section considers the case of three SPs forming coalition. The deployment of the nodes and flow sessions of each SP are given in Fig.~\ref{Network_topology}(a), (b), and (c) while the flow sessions under cooperation are shown in Fig.~\ref{Network_topology}(d). Table \ref{Session} gives the details of the flow session profiles in the three-SP network. 
Since the number of SPs in WMN is three, the core can be presented by \emph{barycentric coordinates} as shown in Fig. \ref{Barycentric}. The shadow area represents the unstable imputations with which the grand coalition would not be formed. With this representation, the relationship of core, dual payoff, and Shapley value is straightforward. As observed, dual payoff and Shapley value both locate in the core area which means they provide payoff allocations that stabilize the grand coalition.

Table \ref{cc0} shows the results from two payoff allocation solutions under different coalition structures. For three-SP network, there are five possible coalition structures denoted by $\omega_{1}-\omega_{5}$. As expected, the grand coalition, represented as $\omega_{5}$, 
maximizes the aggregated payoff, and it is the only stable coalition structure under which each SP gains a higher payoff than that under any other coalition structures. 

\begin{table*}[ht]
\centering
\caption{\footnotesize Payoff Matrix for Three-SP Coalitional Game Without Coalition Cost, } \label{cc0} 
\begin{tabular}{|l|l|l|l|l|l|l|l|} 
\hline
Coalition Structure & $\mu_{1}(v(\mathcal{M}))$ & $\mu_{2}(v(\mathcal{M}))$ & $\mu_{3}(v(\mathcal{M}))$ & $\varphi_{1}(v(\mathcal{M}))$ & $\varphi_{2}(v(\mathcal{M}))$ & $\varphi_{3}(v(\mathcal{M}))$ & $v(\mathcal{M})$ \\ \hline
\hline
$\omega_{1}=\{\{$SP$1\},\{$SP$2\}, \{$SP$3\}\}$ & $767$ & $1101$ & $976$ & $767$ & $1101$ & $976$ & $2844$ \\
$\omega_{2}= \{\{$SP$1$, SP$2\}, \{$SP$3\}\}$  & $800$ & $1101$ & $976$ & $783.5$ & $1117.5$ & $976$ & $2877$ \\
$\omega_{3}=\{\{$SP$1\}, \{$SP$2$, SP$3\}\}$ & $767$ & $1149$ & $1058$ & $767$ & $1166$ & $1041$ & $2974$ \\
$\omega_{4}=\{\{$SP$1$, SP$3\}, \{$SP$2\}\}$ & $822$ & $1101$ & $1013$ & $813$ & $1101$ & $1022$ & $2936$ \\
$\omega_{5}=\{\{$SP$1$, SP$2$, SP$3\}\}$ & $855$ & $1149$ & $1058$ & $817\frac{1}{6}$ & $1170\frac{1}{6}$ & $1074\frac{2}{3}$ & $ 3062$  \\   
\hline           
\end{tabular}
\end{table*}

Taking a close look at the results in Table \ref{cc0}, it can be found that dual payoff solution allocates aggregated payoff to each SP with an amount corresponding to the revenue minus the cost of using the relay nodes. For example, the total flow rate requirement of SP$1$ is $33+42+55=130$Kbps. As this rate requirement can be satisfied, a revenue of $1300$ shall be collected according to the first term of (\ref{lpg}). From Fig. \ref{Network_topology}(d) we can observe that the number of nodes involved in the transmission of flow sessions $l^{1}_{1}$, $l^{1}_{2}$ and $l^{1}_{3}$ are $3$, $3$ and $4$, respectively. The cost calculated according to the second term of (\ref{lpg}) is $33\times 3 + 42\times 3 +55 \times 4 = 445$. The allocated revenue subtracting the cost gives $855$ which equals to the payoff of SP$1$ under dual payoff solution. This is also the case with SP$2$ and SP$3$. 
It is because the dual payoff solution allocates payoff based on the amount of occupied resource, i.e., utilizing nodes in this SPs coalitional game. If an SP uses links supported by other SP's nodes, the cost of those nodes are transferred from latter to former accordingly. However, there is no such relationship in Shapley value solution. As described in Section V, Shapley value solution allocates aggregated payoff according to the marginal contribution of each SP. That is, if a node of an SP is used by another SP, the payoff generated is shared between the two SP according to (\ref{sv}). Each solution provides the most fair solution based on its allocation principle and both solutions are stable in this SPs coalitional game.

\section{Conclusion}
In this paper, we have modeled the payoff optimization problem of SPs cooperation as a coalitional game with transferable utility, specifically, a linear programming game.
Based on the concepts of dual payoff and Shapley value, we have obtained the stable solutions for the formulated linear programming game under general network topology. 
For the future work, the distributed algorithm of payoff allocation solution with coalition formation of SPs will be developed.


\begin{thebibliography}{99}

\bibitem{Lin10}
P. Lin,   J. Jia,   Q. Zhang, and M. Hamdi, ``Cooperation among wireless service providers: opportunity, challenge, and solution", \emph{IEEE Wireless Communications}, Vol. 17, No. 4, pp. 55-61, August 2010.

\bibitem{Guha07}
R. K. Guha, ``Resource sharing and allocation in wireless mesh networks", Dissertations available from ProQuest. Paper AAI3271760.

\bibitem{Driessen10}
S. H. Driessen, ``Cooperative Games, Solutions and Applications", Springer, 1st Edition, November 2010. 


\bibitem{Shapley53}
L. S. Shapley, ``A value for n-person games. in contributions to the theory of games", \emph{Annals of Mathematical Studies}, vol. 28, no. 2, 1953.

\bibitem{Richard10}
T.B. Ma, D. M. Chiu, C.S. Lui, V. Misra and D. Rubenstein, ``Internet Economics: The Use of Shapley Value for ISP Settlement", 
\emph{IEEE/ACM Transactions on Networking}, June 2010. 

\bibitem{Maac}
T.B. Ma, D. M. Chiu, C.S. Lui, V. Misra and D. Rubenstein, ``On Cooperative Settlement Between Content, Transit and Eyeball Internet Service Providers", \emph{IEEE/ACM Transactions on Networking}, Accepted. 

\bibitem{Le08}
V. Le, Z. Feng, P. Zhang, Y. Huang, and X. Wang, ``A Dynamic Spectrum Allocation Scheme with Interference Mitigation in Cooperative Networks", in \emph{Proceeding of IEEE WCNC}, March 2008.

\bibitem{Pan09}
M. Pan, F. Chen, X. Yin, and Y. Fang, ``Fair Profit Allocation in the Spectrum Auction Using the Shapley Value", in \emph{Proceeding of IEEE GLOBECOM}, Honolulu, Hawaii, USA, November 2009.

\bibitem{Niyato06}
D. Niyato and E. Hossain, ``A cooperative game framework for bandwidth allocation in 4G heterogeneous wireless networks", in \emph{Proceedings of IEEE ICC}, Istanbul, Turkey, 11-15 June, 2006. 

\bibitem{winter02}
E. Winter, \emph{The Shapely Value, in the Handbook of Game Theory}. R. J. Aumann and S. Hart, North-Holland, 2001. 


\bibitem{hou08}
Y. T. Hou, Y. Shi, and H. D. Sherali, ``Spectrum Sharing for Multi-Hop Networking with Cognitive Radios", \emph{IEEE Journal on Selected Areas in Communications}, Vol.26, No.1, pp. 146-155, January 2008.

\end{thebibliography}
\end{document}